\documentclass[aps,prb,reprint,superscriptaddress,noeprint]{revtex4-2}

\usepackage{mathtools}
\usepackage{braket}
\usepackage{graphicx}
\usepackage{placeins}
\usepackage{blkarray}
\graphicspath{{figures/}}
\usepackage{color}

\renewcommand{\vec}[1]{\mathbf{#1}}

\begin{document}

\title{Constrained Random Phase Approximation of the effective Coulomb interaction in lattice models of twisted
  bilayer graphene}

\author{Tuomas I. Vanhala}

\author{Lode Pollet}

\affiliation{Department of Physics,
Arnold Sommerfeld Center for Theoretical Physics,
University of Munich, Theresienstr. 37, 80333 M\"{u}nchen, Germany}
\affiliation{Munich Center for Quantum Science and Technology (MCQST),
Schellingstr. 4, 80799 M\"{u}nchen, Germany}

\date{\today}

\begin{abstract}
  Recent experiments on twisted bilayer graphene show the urgent need
  for establishing a low-energy lattice model for the system. We use
  the constrained random phase approximation to study the interaction
  parameters of such models taking into account screening from the
  moire bands left outside the model space. Based on an atomic-scale
  tight-binding model, we develop a numerically tractable
  approximation to the polarization function and study its behavior
  for different twist angles. We find that the polarization has three
  different momentum regimes. For small momenta, the polarization is
  quadratic, leading to a linear dielectric function expected for a
  two-dimensional dielectric material. For large momenta, the
  polarization becomes independent of the twist angle and approaches
  that of uncoupled graphene layers. In the intermediate momentum
  regime, the dependence on the twist-angle is strong. Close to the
  largest magic angle the dielectric function peaks at a momentum of
  $~1/(4 \: nm)$ attaining values of 25, meaning very strong screening
  at intermediate distances. We also calculate the effective screened
  Coulomb interaction in real space and give estimates for the on-site
  and extended interaction terms for the recently developed
  hexagonal-lattice model. For free-standing TBG the effective
  interaction decays slower than $1/r$ at intermediate distances $r$,
  while it remains essentially unscreened at large enough $r$.
\end{abstract}

\maketitle

\section{Introduction}

There is currently no consensus on an effective low-energy model which
can describe the unexpected experimental observation of correlated
insulating states and superconductivity in twisted bilayer graphene
(TBG)
\cite{CaoSuperconductivity,CaoInsulator,Yankowitz1059,2019arXiv190205151C,Sharpeeaaw3780}. The
non-interacting band structure of TBG is modeled by tight-binding
calculations \cite{PhysRevLett.99.256802,PhysRevB.82.121407}, and can
also be calculated from a continuum approximation
\cite{Bistritzer2011}. Close to the charge neutrality point, these
calculations reveal a structure of four (spin-degenerate) bands, which
can become narrower than $10 \: meV$ close to certain ``magic'' twist
angles. The experimentally found insulating states at a doping of
$\pm 4$ electrons per moire unit cell are then explained by band gaps
separating this structure from the rest of the spectrum. However, the
exact bandwidth at a given angle is sensitive to lattice relaxation
effects and the choice of the hopping parameters
\cite{Yoo2019,PhysRevB.99.205134,PhysRevB.98.081410}. Recent electron
microscopy experiments
\cite{Kerelsky_2019,2019arXiv190102997C,PhysRevB.99.201408} seem to
indicate bandwidths of a several tens of $meV$ at the angle of
$\sim 1.1 \deg$ where the transport measurements
\cite{CaoSuperconductivity,CaoInsulator} were performed, while the
narrowest bandwidths are obtained for smaller twist angles than
predicted theoretically \cite{Bistritzer2011}, possibly due to an
interaction-induced renormalization of the single layer Fermi velocity
\cite{Kerelsky_2019}.  Instabilities
on triangular
\cite{PhysRevB.99.195120,PhysRevB.98.075154,PhysRevB.98.085436,PhysRevB.97.235453,PhysRevLett.121.087001}
and hexagonal
\cite{PhysRevLett.122.246401,PhysRevB.98.241407,PhysRevB.99.094521,PhysRevB.98.075109,PhysRevB.98.245103,Pizarro_2019,PhysRevB.98.045103}
low-energy lattice models, as well as the full microscopic
tight-binding model or the corresponding continuum theory
\cite{PhysRevB.99.121407,PhysRevB.98.205151,Younpj2019,LAKSONO201838,acs.nanolett.8b02033,PhysRevLett.122.026801}
have been studied.
The hexagonal lattice model with two spin-degenerate orbitals per site
is theoretically sound, as it can be derived from the maximally
localized Wannier functions of the four low-energy bands
\cite{PhysRevX.8.031087,PhysRevX.8.031088,PhysRevX.8.031089,PhysRevLett.122.246402},
which turn out to have a peculiar three-lobe structure. The triangular
lattice models can then be seen as models of the van Hove
singularities of either the conduction or the valence bands.

To understand the mechanism of the correlated insulating states and
superconductivity, the effective electron-electron interaction needs
to be known, which is a central open question. Some mean-field
treatments approach the superconductivity by postulating
phenomenological attractive interactions
\cite{PhysRevB.98.195101,2018PhRvB..98v0504P,2019arXiv190606313J,2019arXiv190607152H}. The
possibility of a phononic interaction acting as the ``pairing glue''
for the superconductivity has also been discussed
\cite{PhysRevB.98.241412,PhysRevLett.121.257001,PhysRevB.99.195114,PhysRevLett.122.257002}.
Explaining the insulating states in this picture might still require
dominant Coulomb effects at some fillings, although a bosonic Mott
insulator state has also been proposed
\cite{PhysRevLett.122.257002}. Another direction is to start with
repulsive Coulomb interactions. Because of the multitude of different
models, many insulating states have been proposed, ranging from
different magnetic states
\cite{PhysRevB.98.075109,PhysRevB.98.245103}, to spin-
\cite{PhysRevLett.121.217001} and charge-density waves
\cite{PhysRevX.8.041041} and paramagnetic Mott insulators
\cite{PhysRevB.98.235158}, to name a few.

To narrow down the range of possible models, it is important to better
understand the effective interactions within TBG. Here we work towards
this goal by analyzing the Coulomb interactions in low-energy lattice
models. We work under the assumption that correlation effects mainly
happen within the four low-energy bands, and a low-energy lattice
model can thus be built on the corresponding Wannier functions
\cite{PhysRevX.8.031087,PhysRevX.8.031088,PhysRevX.8.031089,PhysRevLett.122.246402}.
Assuming an unknown dielectric constant, the direct Coulomb and
exchange interaction strengths for this model were calculated in
\cite{PhysRevX.8.031087}. However, it is not clear that the dielectric
function can be assumed to be independent of the distance. In fact, it
can be argued on quite general grounds that this is not the case: In a
two-dimensional dielectric the dielectric function is linear in the
wavevector, $\epsilon(\vec{q})=1+2\pi\alpha_{2D}|\vec{q}|$, and the
Coulomb interaction is thus always unscreened at long distances
\cite{PhysRevB.84.085406}. On the other hand, the short range
screening will depend on the details of the specific system. An order
of magnitude estimate is given by the RPA result for eight-component
Dirac Fermions (from spin, layer and valley degeneracy) at the
charge-neutrality point, yielding
$\epsilon=1+8 \cdot 2 \pi k_c/(16 v_F) \approx 10$, where $k_c$ is the
Coulomb constant \cite{RevModPhys.84.1067} and $v_F$ is the graphene
Fermi velocity. One can thus expect rather strong screening for short
distances in TBG, and it is interesting to study how this connects to
the long-range behavior.

Screening of the effective Coulomb interaction that enters the
low-energy model of TBG results from both the environment and from the
graphene itself. Intrinsic screening comes from the bands left outside
the model space, including the graphene $\sigma$-bands
\cite{PhysRevLett.106.236805} as well as the higher moire bands. Here
we concentrate especially on the latter part. We use the microscopic
tight-binding model from \cite{PhysRevB.96.075311} (without the
lattice relaxation) and develop an approximation which allows us to
numerically evaluate the RPA polarization function. As we do not want
to include the polarizability arising from the low-energy subspace
itself, we apply the constrained RPA (cRPA) method
\cite{PhysRevB.70.195104,PhysRevLett.106.236805,PhysRevB.86.165105,PhysRevB.91.245156}
where excitations between the low-energy bands are neglected. We find
that the dielectric function has three different momentum regimes: The
behavior at small momenta is linear, as expected for a two-dimensional
dielectric. The linear approximation remains valid up to a momentum
scale $\sim \Delta/v_F$, where $\Delta$ is related to the energy width
of the low-energy subspace and thus becomes small close to the magic
angle (see Section \ref{PerturbationTheorySection}). In the
intermediate momentum regime the dielectric function attains a maximum
value that depends strongly on the twist, reaching
$\epsilon \approx 25$ close to the magic angle. For momenta
sufficiently larger than $V_{il}/v_F \approx 1/nm$, where $V_{il}$ is
the scale of the local interlayer coupling, the dielectric function
starts to approach that of uncoupled layers and becomes independent of
the twist angle.

The cRPA screening for magic angle TBG has been considered in another
recent paper \cite{2019arXiv190411765P}, where the main focus is on
tuning the effective on-site interaction by changing the dielectric
environment of the TBG. Here we concentrate especially on the extended
interactions, which are found to be strong and long-range in
freestanding TBG. In experimental devices, the importance of the
long-range part depends especially on the gate electrodes, which
provide metallic screening at length scales larger than their distance
from the TBG. Changing this distance makes it possible to increase or
decrease the effective interaction range. We also calculate the
long-range screened interactions in real space, and explicitly
estimate interactions for the hexagonal low-energy lattice model
\cite{PhysRevX.8.031087}. As our results are directly based on the
microscopic tight-binding model \cite{PhysRevB.96.075311}, a similar
treatment might be applicable to other systems for which a simple
continuum theory \cite{Bistritzer2011}, as used in
\cite{2019arXiv190411765P}, is not known.  

\section{Model and method}

\subsection{Polarization function within microscopic tight-binding model}

We model the TBG using the unrelaxed tight binding model from
\cite{PhysRevB.96.075311} and perform the calculations for strictly
periodic moire structures. Using the same notation as in
\cite{PhysRevB.96.075311}, the moire lattice lattice vector can be
written as $m\vec{a}_1+n\vec{a}_2$, where $m$ and $n$ are integers and
$\vec{a}_1$ and $\vec{a}_2$ are the lattice vectors of, say, the upper
graphene layer. The twist angle $\theta$ of the lower layer is then
given by
\begin{equation}
\cos \theta =\frac{1}{2}\frac{m^2+n^2+4mn}{m^2+n^2+mn}.
\end{equation}
We give the parameters of the systems considered in this paper in
Table \ref{system_params_table}. For the present model the first magic
angle, i.e. the largest twist angle where the Fermi velocity is zero,
is $1.20 \: \deg$ \cite{PhysRevB.96.075311}. We also consider several
larger angles, as well as the smaller angle $\theta=1.08 \deg$. The
latter is close to the angle used in the transport experiments
\cite{CaoInsulator,CaoSuperconductivity}, but is below the largest
magic angle within the present model.

\begin{table}
\begin{tabular}{ c | c | c | c | c }
n & m & $\theta$ ($\deg$) & $|\vec{L}_m|$ ($nm$) & $N_a$ \\
\hline
$12$ & $13$ & $2.65$ & $5.32$ & $1876$ \\
$17$ & $18$ & $1.89$ & $7.46$ & $3676$ \\
$22$ & $23$ & $1.47$ & $9.59$ & $6076$ \\
$27$ & $28$ & $1.20$ & $11.72$ & $9076$ \\
$30$ & $31$ & $1.08$ & $12.99$ & $11164$ \\
\end{tabular}
\caption{Parameters of the TBG systems considered in this paper.  The
  indices $m$ and $n$ give one moire lattice vector
  $\vec{L}_m=m\vec{a}_1+n\vec{a}_2$, $\theta$ is the twist angle and
  $N_a$ is the number of atoms in the moire unit cell. Within the
  present model, the magic angle is $\theta=1.20 \deg$.
  \label{system_params_table}}
\end{table}

In general, a tight binding Hamiltonian can be written in the matrix
form
\begin{equation}
H=\sum_{\vec{R}_1,\vec{R}_2} \psi_{\vec{R}_1}^\dagger T_{\vec{R}_1,\vec{R}_2} \psi_{\vec{R}_2},
\end{equation}
where $\psi_{\vec{R}}$ is a vector of annihilation operators in the
unit cell at $\vec{R}$ and $T_{\vec{R}_1,\vec{R}_2}$ is a hopping
matrix. We denote the orbital positions (in our case the carbon atom
positions) within the unit cell by $\vec{a}$ and $\vec{b}$ so that the
matrix element $T_{\vec{R}_1,\vec{R}_2}(\vec{a},\vec{b})$ gives the
tunneling amplitude from orbital $\vec{a}$ in unit cell $\vec{R}_1$
and orbital $\vec{b}$ in unit cell $\vec{R}_2$. All vectors are
two-dimensional: the component perpendicular to the graphene plane is
treated separately.

We can perform a Fourier transformation to find the Bloch Hamiltonian
matrix elements at momentum $\vec{k}$ as
\begin{equation}
T_{\vec{k}}(\vec{a},\vec{b})=\sum_{\vec{R}} \exp(i k \cdot \vec{R}) T_{\vec{0},\vec{R}}(\vec{a},\vec{b}).
\end{equation}
We denote the Bloch eigenstates as $c_{\vec{k},n}(\vec{a})$ with
eigenvalues $\epsilon_{\vec{k},n}$, $n$ being a band index. As our
Bloch orbitals have a real space structure, we can further perform a
unitary transformation
\begin{equation}
d_{\vec{k},n}(\vec{a})=\exp(-i \vec{k} \cdot \vec{a}) c_{\vec{k},n}(\vec{a}).
\end{equation}
The states $d$ are then eigenstates of the transformed Bloch
hamiltonian
\begin{equation}
\begin{split}
T'_{\vec{k}}(\vec{a},\vec{b})=\exp(i \vec{k} \cdot (\vec{b}-\vec{a}))T_{\vec{k}}(\vec{a},\vec{b})=\\
\sum_{\vec{R}} \exp(i k \cdot (\vec{R}+\vec{b}-\vec{a})) T_{\vec{0},\vec{R}}(\vec{a},\vec{b}).
\end{split}
\end{equation}
We do not include spin-flipping terms in the Hamiltonian, and treat
the spin degeneracy implicitly.

Within the RPA the polarization function is given by the bare bubble
diagram, which, for a time-reversal-invariant system at zero
temperature, can be evaluated as
\cite{PhysRevB.70.195104,PhysRevLett.106.236805,PhysRevB.86.165105,PhysRevB.91.245156}
\begin{widetext}
\begin{equation}
\begin{split}
P(\Omega,\vec{q};\vec{a},\vec{b})=\frac{1}{N_k} \sum_{\vec{k},n,n'} c_{\vec{k},n}(\vec{a}) c_{\vec{k}',n'}(\vec{a})^* c_{\vec{k}',n'}(\vec{b}) c_{\vec{k},n}(\vec{b})^*
\left( \frac{1}{\Omega+\epsilon_{\vec{k},n}-\epsilon_{\vec{k}',n'}+i\eta} - \frac{1}{\Omega-\epsilon_{\vec{k},n}+\epsilon_{\vec{k}',n'}-i\eta} \right) \cdot \\
\cdot \delta(\epsilon_{\vec{k},n}<0 \land \epsilon_{\vec{k}',n'} >0),
\end{split}
\label{RPAEquation}
\end{equation}
\end{widetext}
where $\vec{k'}=\vec{k}-\vec{q}$, and we again consider
$P(\Omega,\vec{q})$ to be a matrix in the orbital positions $\vec{a}$
and $\vec{b}$. For notational convenience we assume here that the band
energies $\epsilon_{\vec{k},n}$ are measured with respect to the fermi
energy. From now on we also leave out the imaginary infinitesimals
$i\eta$, which are irrelevant in a gapped system for small
$\Omega$. For the \emph{constrained} RPA the only difference is that
terms of the sum where both bands, $n$ and $n'$, belong to the chosen
subspace are left out of the sum. This is done to avoid double
counting, as the screening effects within this subspace will be
treated by other methods more suitable for strongly correlated
systems.

If the polarization matrix $P$ is known, one can then calculate the
screened interaction $W$ for a given momentum exchange and frequency
as
\begin{equation}
W(\vec{q},\Omega)=\left( I - 2 V(\vec{q}) P(\vec{q},\Omega) \right)^{-1} V(\vec{q}),
\label{cRPAScreenedInteractionEquation}
\end{equation}
wehere $I$ is the identity matrix, $V$ the Coulomb matrix and the
factor of $2$ is included to account for the spin. Formally speaking,
our goal in the following is to find a basis where the matrices $P$
and $V$ are approximately diagonal, so that the matrix inversion
becomes trivial.

\subsection{Coulomb interaction}

\label{CoulombInteractionSection}

For pure graphene the bare Coulomb interaction (i.e. without the
$\sigma$-band screening effects) between the localized Wannier
orbitals of the $\pi$-bands has been reported in
\cite{PhysRevLett.106.236805}. For distances larger than a few lattice
vectors the potential is well approximated by the simple Coulomb form
$V_0(r)=e^2(4\pi \epsilon_0 r)^{-1}$. In fact, comparing to the DFT
results \cite{PhysRevLett.106.236805}, even the nearest neighbour
interaction is given to around $20\%$ accuracy and the next nearest
within $10\%$, when $r$ is taken to be the distance between the carbon
atoms. It is reasonable to assume that this also holds for the
interlayer interactions, while the bare on-site interaction has been
estimated \cite{PhysRevLett.106.236805} as $U=17 eV$. We thus define
the real space potential $V$ so that $V(\vec{b}-\vec{a}) = V_0( r)$
for $ r = \left| \vec{b}-\vec{a} \right| \neq 0$ and $V(0) = U$.

In principle we want to calculate the Fourier transform
\begin{widetext}
\begin{equation}
\tilde{V}^{SL}_{s_1 l_1, s_2 l_2}(\vec{q},\vec{Q}_1,\vec{Q}_2)=\frac{4A}{N_{a} N_{k}} \sum_{\vec{a} \in l_1 s_1,\vec{b} \in l_2 s_2} \exp(-i(\vec{q}+\vec{Q}_1) \cdot \vec{a}) V(\vec{b}-\vec{a}) \exp(i(\vec{q}+\vec{Q}_2) \cdot \vec{b}),
\label{VFourierTransformDefinition1}
\end{equation}
where $\vec{q}$ is a vector in the first Brillouin zone of the
superlattice and $\vec{Q}_1$ and $\vec{Q}_2$ are superlattice
reciprocal lattice vectors. Here the superscript $SL$ refers to a
``sublattice and layer'' basis. The summations over $\vec{a}$ and
$\vec{b}$ are restricted to a given sublattice $s_{1/2} \in \{A,B\}$
and layer $l_{1/2} \in \{0,1\}$, and we consider
$\tilde{V}^{SL}(\vec{q},\vec{Q}_1,\vec{Q}_2)$ to be a $4 \times 4$
matrix in the combined sublattice and layer index. The graphene unit
cell area $A$ is included for comparison to continuum approximations,
and $N_a/4$ is the number of graphene cells in the superlattice unit
cell.

We evaluate the elements
$\tilde{V}^{SL}_{A0,A0}(\vec{q},\vec{Q},\vec{Q})$ and
$\tilde{V}^{SL}_{A0,B0}(\vec{q},\vec{Q},\vec{Q})$ numerically from
Eq. \ref{VFourierTransformDefinition1} and use a momentum-diagonal
approximation of the form
\begin{equation}
\tilde{V}^{SL}(\vec{q},\vec{Q}_1,\vec{Q}_2)=
\begin{bmatrix}
\tilde{V}^{SL}_{A0,A0} & \tilde{V}^{SL}_{A0,B0} & \tilde{V}_{il} & \tilde{V}_{il} \\
(\tilde{V}^{SL}_{A0,B0})^* & \tilde{V}^{SL}_{A0,A0} & \tilde{V}_{il} & \tilde{V}_{il} \\
\tilde{V}_{il} & \tilde{V}_{il} & \tilde{V}^{SL}_{A0,A0} & \tilde{V}^{SL}_{A0,B0}  \\
\tilde{V}_{il} & \tilde{V}_{il} & (\tilde{V}^{SL}_{A0,B0})^* & \tilde{V}^{SL}_{A0,A0}  \\
\end{bmatrix}\delta_{\vec{Q}_1\vec{Q}_2}.
\end{equation}
\end{widetext}
Here we approximate the interlayer part using the simple continuum integral
\begin{equation}
\begin{split}
\tilde{V}_0(\vec{q},d) &= \int d^2\vec{r} \exp(-i \vec{q} \cdot \vec{r}) V_0(\sqrt{r^2+d^2}) \\
&= \frac{2 \pi e^2}{4\pi \epsilon_0 q} \exp(-d q),
\end{split}
\label{LongRangeVEquation}
\end{equation}
so that
$\tilde{V}_{il}(\vec{q},\vec{Q})=\tilde{V}_0(\vec{Q}+\vec{q},d_{il})$,
and $d_{il} \approx 0.334 \: nm$ is the interlayer distance. This
approximation neglects the lattice structure for the interlayer
interaction, while the short distance intralayer terms (including the
on-site term) are still correctly taken into account. Here we have
also neglected the twist of layer $l=1$ by using the intralayer
interaction $\tilde{V}^{SL}_{s_1 0,s_2 0}$ of layer $l=0$, but for
small twist angles this makes little difference.

\begin{figure}[t]
\includegraphics[width=\columnwidth]{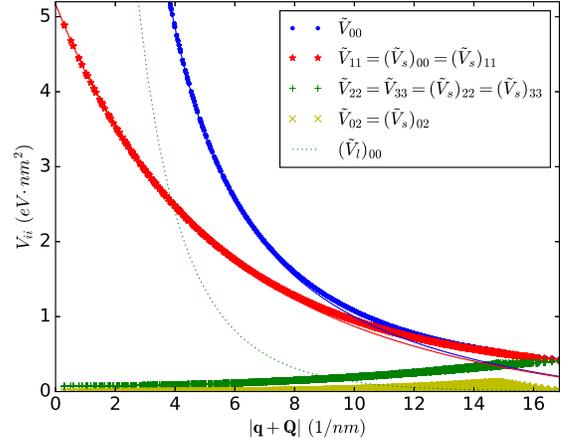}
\caption{Components of the bare Coulomb interaction in momentum space
  within the present approximation. The solid lines represent
  Eqns. \ref{FullV00Approximation} and \ref{FullV11Approximation},
  while the dashed line is the long-range part of the interaction,
  Eqn. \ref{LongRangeInteractionDefinition}. \label{BareCoulombFigure}}
\end{figure}

The interaction becomes approximately diagonal if we apply the
transformation matrix $C_{sl,i}$,
\begin{equation}
C=
\begin{blockarray}{ccccc}
0 & 1 & 2 & 3 \\
\begin{block}{[rrrr]r}
1 &  1 &  1 &  1 & A0 \\
1 &  1 & -1 & -1 & B0 \\
1 & -1 &  1 & -1 & A1 \\
1 & -1 & -1 &  1 & B1 \\
\end{block}
\end{blockarray},
\label{CMatrixDefinitionEqn}
\end{equation}
where the index $i$ indicates the type of a charge distribution, as
explained below. Computing the transformation
$\tilde{V}(\vec{q},\vec{Q}_1,\vec{Q}_2)=\frac{1}{4} C^T
\tilde{V}^{SL}(\vec{q},\vec{Q}_1,\vec{Q}_2) C$,
we obtain the sparse form
\begin{equation}
\tilde{V}(\vec{q},\vec{Q}_1,\vec{Q}_2)=
\begin{bmatrix}
  \tilde{V}_{00} & 0 & \tilde{V}_{02} & 0 \\
  0 & \tilde{V}_{11} & 0 & \tilde{V}_{02} \\
  \tilde{V}_{02}^* & 0 & \tilde{V}_{22} &  0 \\
  0 & \tilde{V}_{02}^* & 0 & \tilde{V}_{22}  \\
\end{bmatrix}\delta_{\vec{Q}_1\vec{Q}_2},
\label{VSparseFormEqn}
\end{equation}
where
\begin{equation}
\begin{split}
\tilde{V}_{00}&=V_{AA}+\Re V_{AB} + 2V_{il} \\
\tilde{V}_{11}&=V_{AA}+\Re V_{AB} - 2V_{il} \\
\tilde{V}_{22}&=V_{AA}-\Re V_{AB}  \\
\tilde{V}_{02}&=-\Im V_{AB}  \\
\end{split}
\label{TildeVComponents}
\end{equation}
The elements of the interaction matrix are plotted in
Fig. \ref{BareCoulombFigure}. One can see that the off-diagonal
element $\tilde{V}_{02}$ is small and the dominant contributions are
$\tilde{V}_{00}$ and $\tilde{V}_{11}$. The component $\tilde{V}_{00}$
is associated with interactions between unpolarized charge
distributions, where the charge density is the same in both layers,
while $\tilde{V}_{11}$ is the interaction between polarized charge
distributions, where the two layers have opposite charge. The
components $2$ and $3$ are then associated with distributions where
the sublattices are oppositely charged.

We can also construct a long-wavelength approximation with
\begin{equation}
\tilde{V}_{00}(\vec{q},\vec{Q},\vec{Q})=2 \left( \tilde{V}_0(|\vec{q}+\vec{Q}|,0)+\tilde{V}_0(|\vec{q}+\vec{Q}|,d_{il}) \right)+u,
\label{FullV00Approximation}
\end{equation}
where the fitted constant $u=-0.8755 \: eV \cdot nm^2$ models the
short range deviations from the Coulomb law. A corresponding
approximation for the interaction between interlayer ``dipoles'' is
given by
\begin{equation}
  \tilde{V}_{11}(\vec{q},\vec{Q},\vec{Q})=2 \left( \tilde{V}_0(|\vec{q}+\vec{Q}|,0)-\tilde{V}_0(|\vec{q}+\vec{Q}|,d_{il}) \right)+u,
\label{FullV11Approximation}
\end{equation}
where $u$ is the same constant, as it mainly originates from the
intralayer interaction. These approximations work well for
$|\vec{q}+\vec{Q}| \lesssim 6$, as can be seen in
Fig. \ref{BareCoulombFigure}. At this level of approximation the
interaction $\tilde{V}_{ij}$ is diagonal in the indices $i,j$ and the
diagonal elements $\tilde{V}_{22}$ and $\tilde{V}_{33}$ are zero.

For the calculation of the dielectric function and the screened
interaction we divide the interaction in long-range and short-range parts
as follows. We write the intralayer part in real space as
\begin{equation}
V(\vec{b}-\vec{a})=V_0(\sqrt{r^2+d_{il}^2})+V_{s}(\vec{b}-\vec{a}),
\label{ShortRangeInteractionDefinition}
\end{equation}
which defines the short range part $V_s$ and includes all contributions beyond
the interlayer form $V_0(\sqrt{r^2+d_{il}^2})$. For large values of $r$, $V_s$ decays as $r^{-3}$.
As the short range
part is purely intralayer, the transformed potential $\tilde{V}_s$ has
the additional symmetry $\tilde{V}_{s,00}=\tilde{V}_{s,11}$ (see
Eq. \ref{TildeVComponents}). The long-range part $\tilde{V}_l$ is not
much affected by the short-range intralayer details, and is
well-approximated by the continuum form
\begin{equation}
(\tilde{V}_l)_{ij}(\vec{q},\vec{Q}_1,\vec{Q}_2)=4\tilde{V}_0(\vec{q}+\vec{Q}_1,d_{il})\delta_{\vec{Q}_1\vec{Q}_2}\delta_{i0}\delta_{j0},
\label{LongRangeInteractionDefinition}
\end{equation}
which includes the convenient cutoff $\exp(-d_{il}q)$, and only the element
$\tilde{V}_{l,00}$ is nonzero. The long- and short-range parts are
also plotted in Fig. \ref{BareCoulombFigure}.

\subsection{Evaluating the matrix elements of P}

The computational cost of calculating the Bloch states and energies of
the system for some $\vec{k}$-grid with $N_k$ points scales as
$O(N_a^3 N_k)$, where $N_a$ is the number of orbitals in the unit
cell. As the Bloch states are needed multiple times during the
calculation of the polarization function, we save them to disk and
read them back to memory as needed. Calculating the polarization
matrix $P(\Omega,\vec{q})$ for \emph{fixed} $\Omega$ and $\vec{q}$
from Eq. \ref{RPAEquation} then scales as $O(N_a^4 N_k)$, where $N_a$
is the number of orbitals in the unit cell. As the moire unit cells of
twisted bilayer graphene close to the first magic angle contain of the
order of $10^4$ atoms, it is not practically possible to evaluate all
matrix elements of $P(\Omega,\vec{q})$. However, selected matrix
elements can be evaluated much faster. Analogously to the interaction
in Eqn. \ref{VFourierTransformDefinition1}, we consider the fourier
transform
\begin{widetext}
\begin{equation}
\tilde{P}^{SL}_{s_1 l_1,s_2 l_2}(\Omega,\vec{q},\vec{Q}_1,\vec{Q}_2)=
\frac{4}{N_a A} \sum_{\vec{a} \in l_1 s_1,\vec{b} \in l_2 s_2} \exp(-i (\vec{q}+\vec{Q}_1) \cdot \vec{a}) P(\Omega,\vec{q};\vec{a},\vec{b}) \exp(i (\vec{q}+\vec{Q}_2) \cdot \vec{b}),
\label{PolarizationFourierTransform1}
\end{equation}
where $\tilde{P}^{SL}(\Omega,\vec{q},\vec{Q}_1,\vec{Q}_2)$ is again a
matrix in the combined sublattice and layer indices, and the summation
is over orbitals in the specific sublattice and layer within the
superlattice unit cell.

We can write $\tilde{P}$ in the basis corresponding to
Eqn. \ref{VSparseFormEqn} by performing the transformation
$\tilde{P}(\Omega,\vec{q},\vec{Q}_1,\vec{Q}_2)=\frac{1}{4}C^T
\tilde{P}^{SL}(\Omega,\vec{q},\vec{Q}_1,\vec{Q}_2) C$.
If we define the notation $C_i(\vec{a})=C_{s(\vec{a})l(\vec{a}),i}$,
where $i$ is again the charge distribution index defined in
Eqn. \ref{CMatrixDefinitionEqn} and $s(\vec{a})$ and $l(\vec{a})$ are
the sublattice and layer of the atom at $\vec{a}$, we can also write
\begin{equation}
\tilde{P}_{i,j}(\Omega,\vec{q},\vec{Q}_1,\vec{Q}_2)=
\frac{1}{N_a A} \sum_{\vec{a},\vec{b}} C_i(\vec{a}) \exp(-i (\vec{q}+\vec{Q}_1) \cdot \vec{a}) P(\Omega,\vec{q};\vec{a},\vec{b}) \exp(i (\vec{q}+\vec{Q}_2) \cdot \vec{b}) C_j(\vec{b}),
\label{PolarizationFourierTransform2}
\end{equation}
where the summation goes over all orbitals in the unit cell. This
Fourier transform is unitary in the sense that one can think of the
factors
$C_{i}(\vec{b}) \exp(i(\vec{q}+\vec{Q}_2) \cdot \vec{b})/\sqrt{N_a}$
as columns in a unitary transformation matrix applied to the matrix
$P(\Omega,\vec{q})$. Like the interaction $\tilde{V}$, the
polarization matrix $\tilde{P}(\Omega,\vec{q},\vec{Q},\vec{Q})$ is
approximately diagonal especially at low momenta, as discussed in
Section \ref{PolarizationFunctionSection}.

The equation for the Fourier transformed polarization function can be
written compactly using the overlap
\begin{equation}
M_{n,n',i}(\vec{q},\vec{Q},\vec{k})=\frac{1}{\sqrt{N_a}} \sum_{\vec{b}} C_i(\vec{b}) \exp(i (\vec{q}+\vec{Q}) \cdot \vec{b}) c_{\vec{k},n}(\vec{b})^* c_{\vec{k}-\vec{q},n'}(\vec{b})=\frac{1}{\sqrt{N_a}} \sum_{\vec{b}} C_i(\vec{b}) \exp(i \vec{Q} \cdot \vec{b}) d_{\vec{k},n}(\vec{b})^* d_{\vec{k}-\vec{q},n'}(\vec{b}),
\label{GeneralMFormula}
\end{equation}
We then have
\begin{equation}
\begin{split}
\tilde{P}_{ij}(\Omega,\vec{q},\vec{Q}_1,\vec{Q}_2)=\frac{1}{N_k A} \sum_{\vec{k},n,n'} M_{n,n',i}(\vec{q},\vec{Q}_1,\vec{k})^* M_{n,n',j}(\vec{q},\vec{Q}_2,\vec{k})
\left( \frac{1}{\Omega+\epsilon_{\vec{k},n}-\epsilon_{\vec{k}',n'}} - \frac{1}{\Omega-\epsilon_{\vec{k},n}+\epsilon_{\vec{k}',n'}} \right) \cdot \\
\cdot \delta(\epsilon_{\vec{k},n}<0 \land \epsilon_{\vec{k}',n'} >0).
\end{split}
\label{FourierTransformedRPAFormula}
\end{equation}
\end{widetext}
Naively evaluating the $M$-factors for fixed $\vec{q}$ scales as
$O(N_Q N_a^3 N_k)$, where $N_Q$ is the number of superlattice
momenta. The computation can also be performed as an optimized matrix
product by considering the matrices
$D^l(i,\vec{q},\vec{Q},\vec{k})_{n,\vec{b}}=C_i(\vec{b}) \exp(i
\vec{Q} \cdot \vec{b}) d_{\vec{k},n}(\vec{b})^*$
and
$D^r(i,\vec{q},\vec{Q},\vec{k})_{\vec{b},n}=d_{\vec{k}-\vec{q},n}(\vec{b})$,
so that
$M_i(\vec{q},\vec{Q},\vec{k})=D^l(i,\vec{q},\vec{Q},\vec{k})
D^r(i,\vec{q},\vec{Q},\vec{k})/\sqrt{N_a}$,
which is considered as a matrix equation in the $n$ and $\vec{b}$
indices.

We can calculate a momentum-space submatrix of $\tilde{P}$ by
selecting only some momenta $\vec{Q}$. For example, if we only want
the momentum-diagonal element, then $N_Q=1$. Once the $M$-factors are
known, computing $\tilde{P}(\Omega,\vec{q})$ from
Eq. \ref{FourierTransformedRPAFormula} scales as $O(N_Q^2 N_a^2 N_k)$,
and is subdominant for large enough $N_a$, if $N_Q$ is kept
constant. If we take the full momentum basis, then $N_Q=N_a/4$ and we
regain the $O(N_a^4 N_k)$ scaling.

\subsection{Perturbation theory at $q \rightarrow 0$}

\label{PerturbationTheorySection}

For a dieletric where the Fermi energy lies in a gap, the relevant
component of the polarization function goes to zero quadratically as
$q \rightarrow 0$. This long wavelength component $P_l$ can be defined
as
$P_l(\Omega,\vec{q})=\tilde{P}_{00}(\Omega,\vec{q},\vec{0},\vec{0})$.
We then only need a simplified M-factor that is given by
\begin{equation}
M_{n,n'}(\vec{q},\vec{k})=M_{n,n',0}(\vec{q},\vec{Q}=0,\vec{k})=\frac{1}{\sqrt{N_a}}\braket{d_{\vec{k},n}|d_{\vec{k}',n'}}.
\end{equation}
From the orthogonality of the eigenstates we find that
$M_{n,n'}(\vec{q}=0,\vec{k})=\frac{\delta_{n,n'}}{\sqrt{N_a}}$. The
next order term is given by perturbation theory as
\begin{equation}
M_{n,n'}(\vec{q},\vec{k})=\frac{\delta_{n,n'}}{\sqrt{N_a}}+\frac{1}{\sqrt{N_a}} \frac{ \braket{ d_{\vec{k},n} | -\frac{\partial T'_{\vec{k}}}{\partial \vec{k}} | d_{\vec{k},n'} } }{\epsilon_{\vec{k},n'}-\epsilon_{\vec{k},n}} \cdot \vec{q} + O(q^2)
\label{MExpansion}
\end{equation}
It should be noted that this equation does not hold for degenerate
bands $n$ and $n'$. However, the degenerate case is never actually
needed when the Fermi level is in a gap, or when we calculate the
constrained polarization and the Fermi level is within the strongly
correlated subspace.

Using the overlap $M$ the long range polarization function can be
written as
\begin{equation}
\begin{split}
P_l(\Omega,\vec{q})=\frac{1}{N_k A} \sum_{\vec{k},n,n'} M_{n,n'}(\vec{q},\vec{k})^* M_{n,n'}(\vec{q},\vec{k}) \cdot \\
\cdot \left( \frac{1}{\Omega+\epsilon_{\vec{k},n}-\epsilon_{\vec{k}',n'}} - \frac{1}{\Omega-\epsilon_{\vec{k},n}+\epsilon_{\vec{k}',n'}} \right) \cdot \\
\cdot \delta(\epsilon_{\vec{k},n}<0 \land \epsilon_{\vec{k}',n'} >0).
\end{split}
\label{RPAEquation2}
\end{equation}
From Eq. \ref{MExpansion} it then follows that $P_l(\Omega,\vec{q}=0)=0$, and 
\begin{equation}
\left. \frac{\partial P_l(\Omega,\vec{q})}{\partial q_i}\right|_{\vec{q}=0}=0.
\end{equation}
Taking the Jacobian, we get
\begin{widetext}
\begin{equation}
\begin{split}
\left. \frac{\partial^2 P_l(\Omega,\vec{q})}{\partial q_i \partial q_j} \right|_{\vec{q}=0} = 
\frac{1}{N_k N_a A} \sum_{\vec{k},n,n'} \left( 
\frac{ \braket{ d_{\vec{k},n'} | \frac{\partial T'_{\vec{k}}}{\partial k_i} | d_{\vec{k},n} } }{\epsilon_{\vec{k},n'}-\epsilon_{\vec{k},n}} 
\frac{ \braket{ d_{\vec{k},n} | \frac{\partial T'_{\vec{k}}}{\partial k_j} | d_{\vec{k},n'} } }{\epsilon_{\vec{k},n'}-\epsilon_{\vec{k},n}}
+ c.c. \right) \\
\cdot \left( \frac{1}{\Omega+\epsilon_{\vec{k},n}-\epsilon_{\vec{k},n'}} - \frac{1}{\Omega-\epsilon_{\vec{k},n}+\epsilon_{\vec{k},n'}} \right)
\delta(\epsilon_{\vec{k},n}<0 \land \epsilon_{\vec{k},n'} >0).
\end{split}
\label{QuadraticPolarizationFormula}
\end{equation}
For the static case $\Omega=0$ this simplifies to
\begin{equation}
\left. \frac{\partial^2 P_l(\Omega=0,\vec{q})}{\partial q_i \partial q_j} \right|_{\vec{q}=0} = 
\frac{-2}{N_k N_a A} \sum_{\vec{k},n,n'} \left( 
\frac{ \braket{ d_{\vec{k},n'} | \frac{\partial T'_{\vec{k}}}{\partial k_i} | d_{\vec{k},n} } \braket{ d_{\vec{k},n} | \frac{\partial T'_{\vec{k}}}{\partial k_j} | d_{\vec{k},n'} } + c.c. }{(\epsilon_{\vec{k},n'}-\epsilon_{\vec{k},n})^3}
\right)
\delta(\epsilon_{\vec{k},n}<0 \land \epsilon_{\vec{k},n'} >0).
\end{equation}
\end{widetext}

One can estimate the region where the quadratic approximation holds by
estimating the characteristic scale of the current operator as
$\frac{\partial T'_{\vec{k}}}{\partial k_i} \sim v_F$, where $v_F$ is
the Fermi velocity of a single graphene layer,
$v_F \approx 0.52 eV nm$ for our model. The perturbation theory is
then expected to be valid as long as $v_F q \ll \Delta$. For the full
RPA $\Delta$ would be the gap between the subspaces above and below
the Fermi level, but for the cRPA it is the gap from the highest state
below the Fermi level to the lowest state above the Fermi level that
is not within the chosen correlated subspace. Close to the first magic
angle the cRPA gap is of the order of $10 meV$, which gives a momentum
scale of $0.02/nm$. Thus the polarization function should cross over
to the dielectric behavior at a relatively large length scale of the
order of $50 \: nm$. However, the actual gap at
$\theta \approx 1.1 \: \deg$ is experimentally seen to be closer to
$50 \: meV$, which is consistent with renormalization of the single
layer Fermi velocity by interactions \cite{Kerelsky_2019}, and thus
the 2D dielectric behavior might extend to much shorter length scales
of $~10 \: nm$.

\section{Results}

\subsection{Polarization function}

\label{PolarizationFunctionSection}

\begin{figure}[t]
\includegraphics[width=\columnwidth]{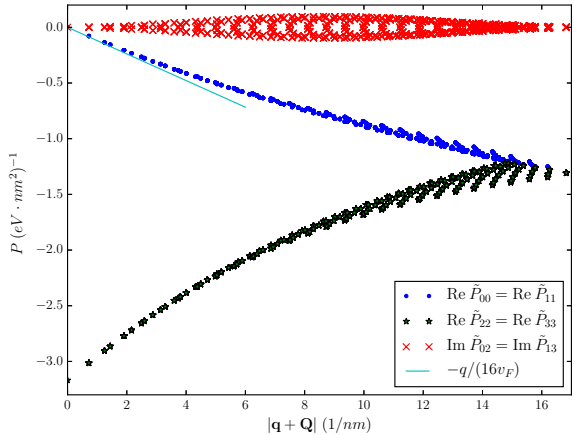}
\caption{Polarization function of uncoupled graphene layers as a
  function of the momentum magnitude. The spreading of the points at a
  given momentum is because the polarization function is not exactly
  isotropic, especially at larger
  momenta. \label{SingleLayerPolarizationFigure}}
\end{figure}

\begin{figure*}[t]
\includegraphics[width=2.0\columnwidth]{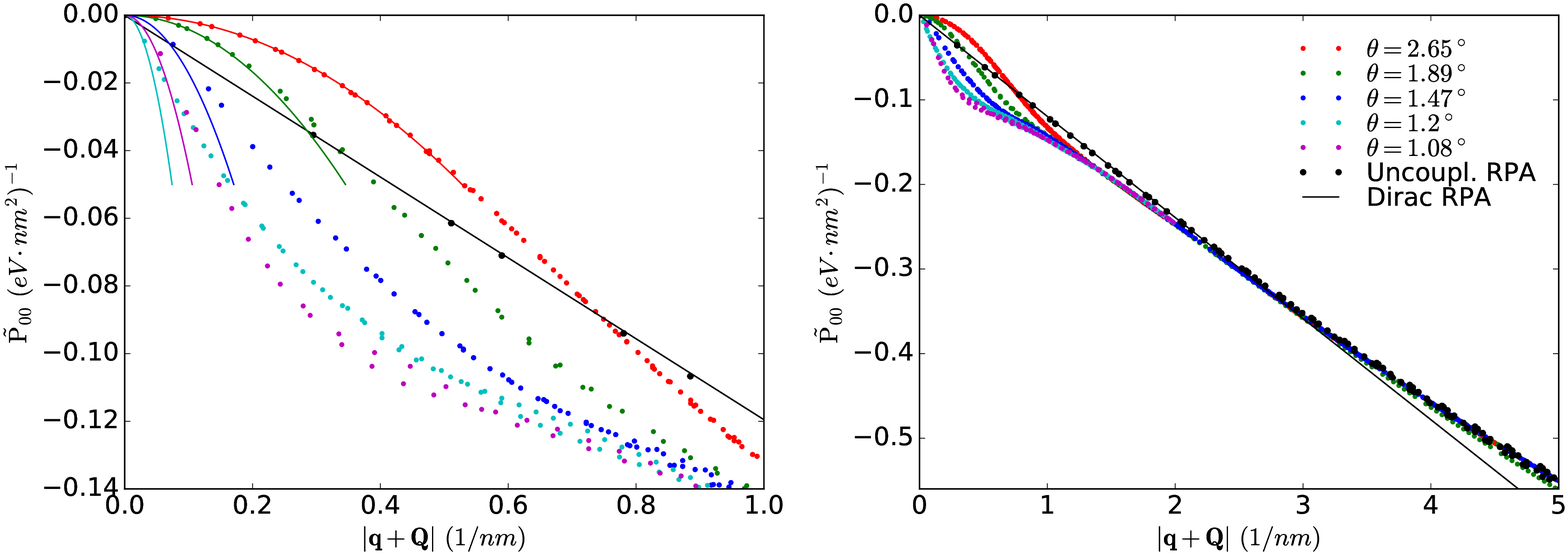}
\caption{Component $\tilde{P}_{00}$ of the polarization function for
  different twist angles. Uncoupled RPA refers to the full RPA result
  for untwisted, uncoupled layers, while Dirac RPA is the analytical
  result for the Dirac approximation of graphene. The solid curves in
  the left panel represent the long-wavelength quadratic approximation
  $\tilde{P}_{00}=-Cq^2$, where $C$ is given in table
  \ref{quadratic_constants_table}. The right panel shows how the
  polarization approaches the case of uncoupled layers for large
  momenta. \label{Polarization0Figure}}
\end{figure*}

In this section we present the results for the cRPA polarization
function for the TBG $\pi$-band electrons in the static case,
i.e. $\Omega=0$. The discussion of the dielectric function and the
screening from the $\sigma$-bands and the environment is left to the
next section. To better understand the effect of the interlayer
coupling, we first briefly discuss the full RPA result for uncoupled,
untwisted layers which is plotted in
Fig. \ref{SingleLayerPolarizationFigure}. In this case the
polarization is fully momentum diagonal, and has the same form as the
short-range part of the interaction,
\begin{equation}
  \tilde{P}=
\begin{bmatrix}
\tilde{P}_{00} & 0 & \tilde{P}_{02} & 0 \\
0 & \tilde{P}_{00} & 0 & \tilde{P}_{02} \\
\tilde{P}_{02}^* & 0 & \tilde{P}_{22} & 0 \\
0 & \tilde{P}_{02}^* & 0 & \tilde{P}_{22} \\
\end{bmatrix},
\end{equation}
where the off-diagonal elements are purely imaginary, and diagonal
elements are real so that the whole matrix is hermitian. For small
momenta the polarization function $\tilde{P}_{ij}$ is nearly diagonal
in the indices $i,j$, and has two branches. The branch
$\tilde{P}_{00}=\tilde{P}_{11}$ approaches zero linearly as
$\tilde{P}_{00}(q)=-q/(16v_F)$, which can be derived using the Dirac
Hamiltonian approximation of graphene
\cite{RevModPhys.84.1067}\footnote{In our notation the polarization
  function corresponds to that of a single Dirac cone, while the
  interaction in the long range limit (see Eq.\
  \ref{LongRangeVEquation}) gets a factor of 4 corresponding to two
  sublattices and two layers.}. This branch is associated with charge
distributions that are equally distributed between the graphene
sublattices. The branch $\tilde{P}_{22}=\tilde{P}_{33}$ is then
associated with oppositely charged sublattices, and approaches a
constant value for $q \rightarrow 0$. For large enough momenta the
polarization starts to deviate from the Dirac approximation and also
gains some off-diagonal elements.

We now consider the full tight-binding model of TBG. As for the
interaction, we consider an approximation where
$\tilde{P}_{i,j}(\Omega,\vec{q},\vec{Q}_1,\vec{Q}_2)$ is diagonal in
the moire reciprocal lattice momenta $\vec{Q}_1$ and $\vec{Q}_2$. We
first diagonalize the Bloch Hamiltonian on a momentum grid storing the
Bloch states on hard disk, and then evaluate Eq.
\ref{FourierTransformedRPAFormula} using the cached results. The grid
sizes used were $20 \times 20$ moire unit cells for $n=12$ and $n=17$
and $n=27$ and $10 \times 10$ for $n=22$ and $n=30$.

Let us first discuss the long-wavelength behavior of the
polarization, starting with the component $\tilde{P}_{00}$, which is
plotted in Fig. \ref{Polarization0Figure}. This is the most relevant
component of the polarization at long wavelengths, as it couples to
the divergent $1/q$-part of the potential. For small enough momentum
exchange the polarization becomes quadratic in $q$,
$P_{00}(\vec{q})=-C q^2$, as discussed in Section
\ref{PerturbationTheorySection}. The results for $C$ are presented in
table \ref{quadratic_constants_table}. For the smaller unit cells
($n=12$ and $n=17$) we have evaluated the coefficient using numerical
integration of Eq.  \ref{QuadraticPolarizationFormula} over
$\vec{k}$. As the numerical integration procedure requires thousands
of integrand evaluations, we cannot perform this procedure for the
larger unit cells, and instead evaluate
Eq. \ref{QuadraticPolarizationFormula} on the same momentum grid as
the full polarization function, yielding less accurate estimates. The
coefficient $C$ can be converted to the $2D$ polarizability
$\alpha_{2D}$ by multiplying with the constant $8 k_c e^2$, where
$k_c=(4 \pi \epsilon_0)^{-1}$ is the Coulomb constant and the factor
of $8$ takes into account layer, sublattice and spin degeneracies. The
polarizability $\alpha_{2D}$ determines the crossover scale
$r_0=2\pi \alpha_{2D}$ such that the interaction is unscreened for
$r \gg r_0$ \cite{PhysRevB.84.085406}.

\begin{table}[h]
\begin{tabular}{ c | c | c | c | c }
$\theta$ ($\deg$) & $C$ ($1/eV$) & $\alpha_{2D}$ $(nm)$ & $r_0$ $(nm)$ \\
\hline
$2.65$ & $0.1796 \pm 0.0002$ & $2.1$ & $13$ \\
$1.89$ & $0.419 \pm 0.002$ & $4.8$ & $30$ \\
$1.47$ & $1.7$ & $20$ & $124$ \\
$1.20$ & $9.1$ & $105$ & $659$ \\
$1.08$ & $4.5$ & $52$ & $325$ \\
\end{tabular}
\caption{The coefficient $C$ of the quadratic part of the polarization
$P_{00}$ and the corresponding 2D polarization length. The length $r_0$ determines
a scale such that the interaction in freestanding TBG is unscreened for $r \gg r_0$.
The long-range screening is strongest at the magic angle $\theta=1.20 \deg$,
and decreases when the angle is either increased or decreased.
\label{quadratic_constants_table}}
\end{table}

\begin{figure}[t]
\includegraphics[width=\columnwidth]{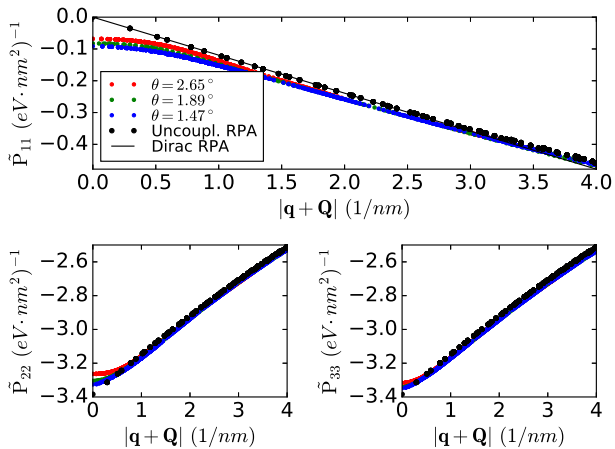}
\caption{Components $\tilde{P}_{11}$, $\tilde{P}_{22}$ and
  $\tilde{P}_{33}$ of the polarization function for different twist
  angles. Uncoupled RPA refers to the full RPA result for untwisted,
  uncoupled layers, while Dirac RPA is the analytical result for the
  Dirac approximation of graphene. \label{Polarization123Figure}}
\end{figure}

In contrast to the component $\tilde{P}_{00}$, the other components do
not have a fixed value at $q \rightarrow 0$. For the components
$\tilde{P}_{22}$ and $\tilde{P}_{33}$ the deviation from the RPA
result of uncoupled layers is not very important, as the shift of
$\tilde{P}_{22}(q=0)$ and $\tilde{P}_{33}(q=0)$ is small relative to
the large absolute value of these components. For $\tilde{P}_{11}$ the
effect of the interlayer coupling is more significant, as it lifts the
degeneracy between the components $\tilde{P}_{00}$ and
$\tilde{P}_{11}$ and makes $\tilde{P}_{11}(q=0)$ non-zero.

For large enough momenta all diagonal components of the polarization
function become independent of the twist angle, and approach the
polarization of uncoupled layers. This can be understood in a real
space perspective: As the coupling between layers is weak relative to
the intralayer couplings, the Green's function for short distances is
governed by the intralayer Hamiltonian, and only over longer distances
the interlayer scattering processes have sufficient time to affect the
propagation. The characteristic time scale of interlayer processes is
proportional to $1/V_{il}$, where $V_{il}=0.48 eV$ is the hopping
amplitude between atoms in the upper and lower layer directly on top
of each other, while the characteristic speed of propagation is the
Fermi velocity $v_F \approx 0.52 \: eV \cdot nm$. This gives a length
scale of $v_F/V_{il} \approx 1\:nm$. For momenta larger than a few
times $1/nm$ the polarization is thus expected to be perturbative in
$V_{il}$ and not depend on the twist angle, as the average local
environment for all twist angles is similar.

In the intermediate momentum regime, where neither the quadratic
approximation nor the approximation of uncoupled layers holds, the
polarization depends strongly on the twist angle. Close to the magic
angle, $|\tilde{P}_{00}|$ is found to be significantly larger than for
uncoupled layers. It is interesting to note that the existence of this
intermediate region is related to the two independent energy scales:
The hopping parameter $V_{il}$ governing the \emph{local} effect of
the interlayer coupling, and the cRPA gap $\Delta$ governing the
low-energy, small-$q$ details. In contrast, in most scenarios of
gapped graphene the gap $\Delta$ has essentially the same magnitude as
the local perturbation, and the crossing from the quadratic behavior
to the short range regime happens directly at the momentum scale
$\Delta/v_F$ with a simpler structure
\cite{PhysRevB.86.195424,Pyatkovskiy_2008}.

We have been unable to find a simple model with a few parameters that
would satisfactorily fit the polarization function for different
angles at all momenta. The model proposed in
\cite{2019arXiv190411765P} (Eqn. (4)) for the magic angle case
reproduces the large polarization in the intermediate region and the
asymptotic Dirac behaviour at large momenta, although we find that the
quantitative fit for our magic angle data is not as good as in the
case of reference \cite{2019arXiv190411765P}. If it also becomes
necessary to model the small-momentum quadratic behaviour, which is
not important for the on-site interaction at the magic angle discussed
in \cite{2019arXiv190411765P}, then a more complicated model with the
two crossover scales discussed above is needed. This is important to
consider, as the width of the low-energy subspace at
$\theta \approx 1.1 \: \deg$ is experimentally found to be wider than
in theoretical magic-angle models
\cite{Kerelsky_2019,2019arXiv190102997C,PhysRevB.99.201408}, and thus
the gap $\Delta$ is also larger, and the quadratic behaviour extends
to larger momenta.

\subsection{Dielectric function and the long-range interaction}

\label{DielectricFunctionSection}

\begin{figure}[t]
\includegraphics[width=\columnwidth]{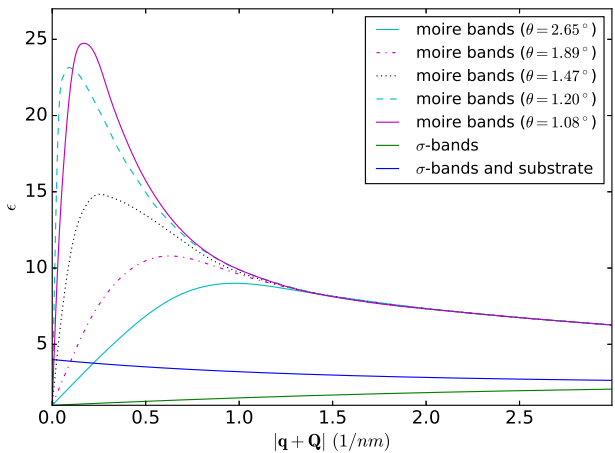}
\caption{Comparison of the long-range component $\epsilon_{00}$ of the
  moire band dielectric function without background screening and the
  background dielectric function $e_{bg}$ with and without the
  substrate contribution. \label{EpsilonLongRangeFigure}}
\end{figure}

Within the cRPA the screened interaction is calculated from
Eq. \ref{cRPAScreenedInteractionEquation} with the interaction and
polarization function discussed in the previous sections. Taking into
account background screening from other sources, such as the graphene
$\sigma$-bands \cite{PhysRevLett.106.236805}, the total dielectric
function can be expressed as
\begin{equation}
\epsilon(\vec{q})=\epsilon_{bg}(\vec{q})-V(\vec{q})P(\vec{q}),
\end{equation}
where $V$ and $P$ are taken to be $4 \times 4$ matrices. We will first
discuss the long-range part of the interaction. For the long-range
part we assume that the polarization and interaction are diagonal, so
that the dielectric function relevant to the long-range interaction is
simply
$\epsilon(\vec{q})=\epsilon_{bg}(\vec{q})-V_{00}(\vec{q})P_{00}(\vec{q})$. We
then use Eq. \ref{FullV00Approximation} to approximate the
interaction $V_{00}$ and a smoothed spline fitted to the polarization
data in Fig. \ref{Polarization0Figure} to represent the polarization
$P_{00}$. The resulting dielectric function with $\epsilon_{bg}=1$, is
shown in Fig. \ref{EpsilonLongRangeFigure}.

\begin{figure*}[p]
\includegraphics[width=1.9\columnwidth]{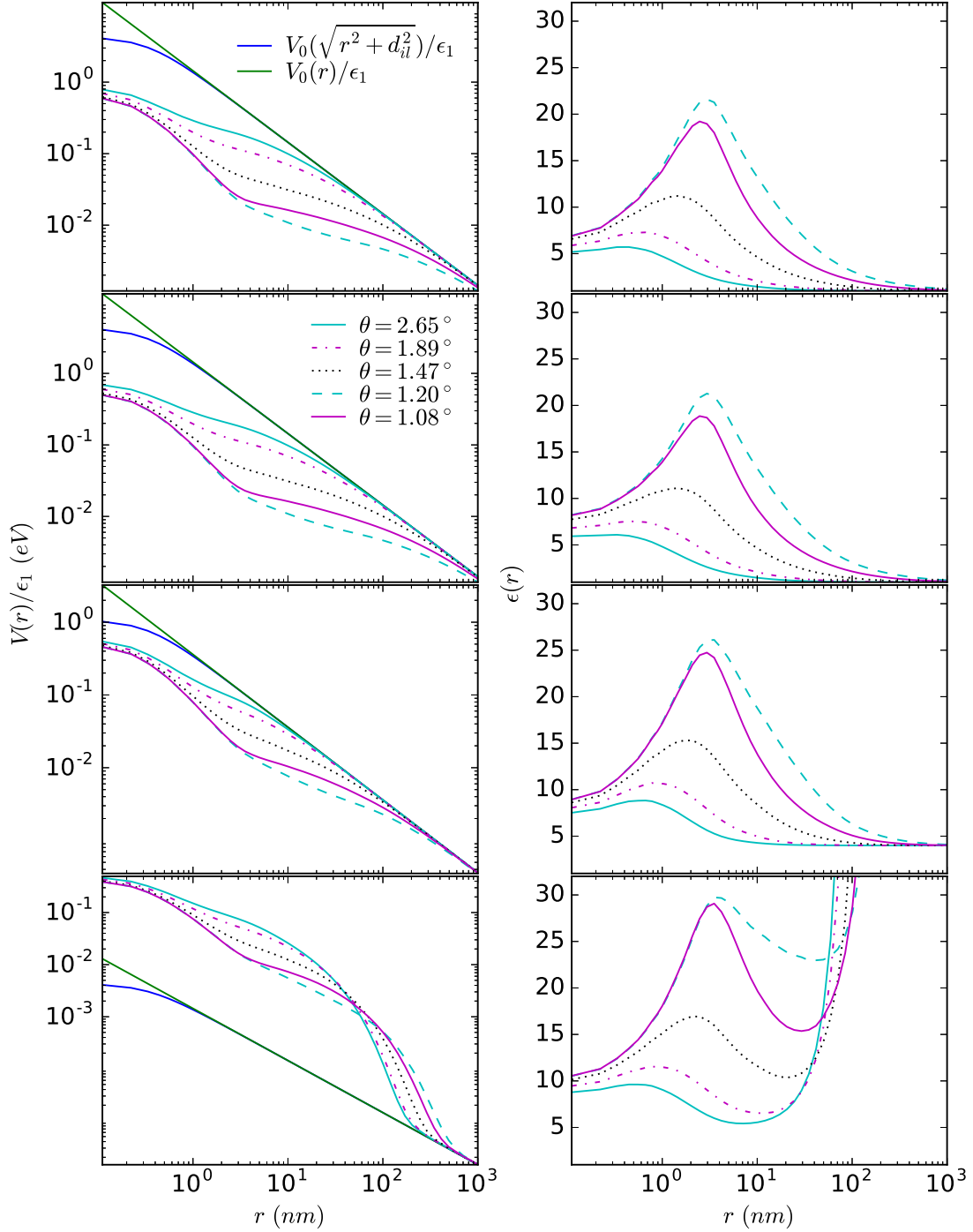}
\caption{The left panels show the screened long-range part of the
  coulomb interaction for different twist angles compared to the
  unscreened interaction, while the right panels show the real-space
  dielectric function defined as
  $\epsilon(r)=V_0(\sqrt{r^2+d^2})/W(r)$, where $W(r)$ is the screened
  interaction. The top row includes screening only from the moire
  bands, the second highest row from moire bands and graphene
  $\sigma$-bands, and the second lowest row also from the boron
  nitride substrate. The lowest row shows how a metallic screening
  from an electrode $30 \: nm$ away from the TBG cuts off the
  interaction in real space. 
  \label{WRLongRangeFigure}}
\end{figure*}

To model the screening sources other than the moire bands we use a
model where a classical dielectric layer of thickness $D$ and
dielectric constant $\epsilon_2$ is embedded in a dielectric
environment with dielectric constant $\epsilon_1$. Following
\cite{Emelyanenko_2008} we evaluate the potential in the middle of the
layer and calculate the effective dielectric function to be
\begin{equation}
\epsilon_{bg}(q)=\epsilon_2 \frac{\epsilon_1\cosh(q D/2)+\epsilon_2\sinh(q D/2)}{\epsilon_2\cosh(q D/2)+\epsilon_1\sinh(q D/2)},
\label{SingleDielectricLayerEpsilon}
\end{equation}
A special case of this model with $\epsilon_1=1$ was used in
\cite{PhysRevLett.106.236805} as an approximation to screening from
the $\sigma$-bands of single layer graphene, which we expect to be the
main additional source of screening also in free-standing TBG. For the
single layer case the parameters are $\epsilon_2=2.4$ and
$D=D_{graphene}=0.28\:nm$. Expanding Eq. \ref{SingleDielectricLayerEpsilon}
in powers of $q$ gives
\begin{equation}
\epsilon_{bg}(q)=\epsilon_1+D q \frac{\epsilon_2^2-\epsilon_1^2}{2\epsilon_2}+O(q^2)=\epsilon_1+2\pi \alpha_{2D,\sigma}q+O(q^2).
\label{ClassicalDielectricLayerModel}
\end{equation}
The effective 2D screening constant for the single layer case is then
$\alpha^{graphene}_{2D,\sigma} \approx 0.044\:nm$. For TBG this value
is expected to double to $\alpha^{TBG}_{2D,\sigma} \approx 0.088\:nm$,
as the $\sigma$-band screening now comes from both layers. Comparing
this value to table \ref{quadratic_constants_table}, we see that the
$\sigma$-bands are insignificant for the long-range screening in
comparison to the moire bands.

To fully treat the local and intermediate-range effects of the
$\sigma$-bonding orbitals would require DFT calculations for the
bilayer system, which is beyond the scope of this work. However, it is
reasonable to assume that the short range screening constant is
roughly the same as for the single layer case. To estimate the overall
importance of the $\sigma$-bands we then use
Eq. \ref{SingleDielectricLayerEpsilon} and set $D=D_{TBG}=0.56\:nm$,
i.e. twice the value found in \cite{PhysRevLett.106.236805}, which
correctly produces the long-distance and short-distance limits
discussed above. The true $\epsilon_{\sigma}$ is expected to have a
more complicated structure, but the relevant length scales
$D_{graphene}$, $D_{TBG}$ and the interlayer distance
$d_{il} \approx 0.334\:nm$ are of similar magnitude, so that the
crossover to the large-momentum behaviour is expected to happen at a
scale of $2...4 \: / nm$.  The estimated contribution of the
$\sigma$-bands to the dielectric function is plotted in
Fig. \ref{EpsilonLongRangeFigure}.

In real experiments the TBG is not free-standing, but typically
sandwiched between boron nitride layers with thickness of $10\:nm$ to
$30\:nm$ \cite{CaoInsulator}. This can be at least qualitatively
modeled by setting $\epsilon_1 \approx 4$
\cite{katsnelson2012graphene}, leading to a background screening with
a negative $\alpha_{2D}$. This background screening model is also
shown in Fig. \ref{EpsilonLongRangeFigure}. On top of the boron
nitride layers the experimental devices also contain metallic gate
electrodes, which are used e.g. to tune the fermi level in the
system. This will provide metallic screening
\cite{2019arXiv190411765P}, which cuts off the interaction at the
momentum scale $1/D_{gate}$, where
$D_{gate} \approx 10 \: nm ... 30 \: nm$ is the distance of the
electrode from the TBG. The corresponding background dielectric
function (not shown) can be modeled qualitatively by neglecting the
$\sigma$-band screening and setting $D=2 D_{gate} \approx 60 \: nm$,
$\epsilon_1=1000$ and $\epsilon_2=4.0$, similarly to the model in
\cite{2019arXiv190411765P}. In the following we will calculate the
interaction in real space, and show the effect of the different
background dielectric functions.

We numerically calculate the Fourier transform of the screened
long-range part of the interaction defined in equation
\ref{LongRangeInteractionDefinition}. This leads to the integral
\begin{equation}
W(r)=\int \frac{d^2 \mathbf{q}}{(2\pi)^2} \exp(i \mathbf{r} \cdot \mathbf{q}) \tilde{V}_{0}(\mathbf{q},d_{il})/\epsilon_{00}(\mathbf{q}),
\end{equation}
where $\tilde{V}_{0}$ is defined in equation
\ref{LongRangeVEquation}. Here the factor $\exp(-q d_{il})$ included
in $\tilde{V}_{0}$ provides a convenient cutoff with the scale
$1/d_{il} \approx 3/nm$. We plot the screened real-space interaction
in the left column of Fig. \ref{WRLongRangeFigure}, and the real-space
dielectric function defined as $\epsilon(r)=V_0(\sqrt{r^2+d^2})/W(r)$
in the right column. In the long-range limit the moire bands do not
provide any screening, so that
$\epsilon(r \rightarrow \infty)=\epsilon_{bg}(q=0)=\epsilon_1$. The
peak in the momentum-space dielectric function
(Fig. \ref{EpsilonLongRangeFigure}), which close to the magic angle is
found at $q \sim 0.25/nm$, translates into a real space peak at
$r \sim 4 \: nm$. Comparing the two upper rows in
Fig. \ref{WRLongRangeFigure}, the $\sigma$-band screening is found to
be rather insignificant. When the boron nitride substrate is taken
into account (third row of Fig. \ref{WRLongRangeFigure}), the peak
value of the dielectric function at the magic angle reaches
$\epsilon \sim 25$. In this model the boron nitride layers are
infinitely thick, so that the screening continues to long
distances. Finally, the lowest panels demonstrate how the interaction
is cut off by screening from metallic electrodes $30 \: nm$ away from
the TBG.

\begin{figure}[t]
\includegraphics[width=\columnwidth]{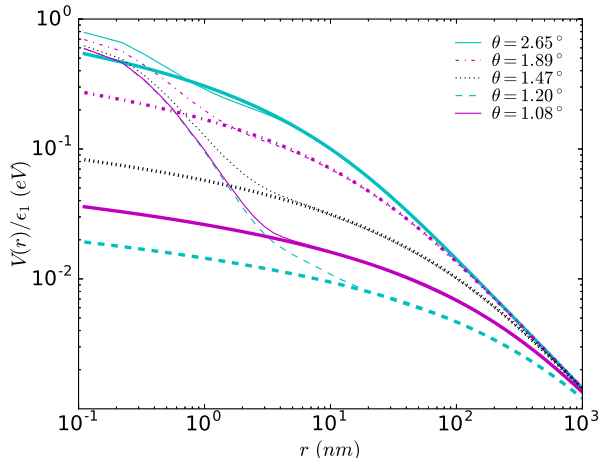}
\caption{Comparison of the long-range component of the screened
  Coulomb interaction (thin lines) to the 2D dielectric screening
  form, Eq. \ref{ScreenedCoulomb2D} (thick lines). Here we have used
  the interaction without the screening from the $\sigma$-bands and
  the environment. \label{Coulomb2DComparisonFigure}}
\end{figure}

The Coulomb interaction screened by a classical two-dimensional
dielectric layer with polarizability $\alpha_{2D}$ can also be
calculated analytically, yielding \cite{PhysRevB.84.085406}
\begin{equation}
  W_{2D}(r)=k_c\frac{e^2}{4\alpha_{2D}}\left( H_0(r/r_0) - Y_0(r/r_0) \right),
  \label{ScreenedCoulomb2D}
\end{equation}
where the $r_0=2\pi \alpha_{2D}$, $k_c$ is the Coulomb constant, $H_0$
is a Struve function and $Y_0$ is a Bessel function of the second
kind. This is a good approximation to the screened interaction of
freestanding TBG beyond a length scale that is essentially determined
by the cRPA gap as discussed in section
\ref{PerturbationTheorySection}. We plot a comparison of this
approximation and the full screened Coulomb interaction in
Fig. \ref{Coulomb2DComparisonFigure}. The screening effect is
well-approximated beyond a few tens of nanometers at the magic angle,
and the approximation improves when moving away from the magic angle.

\subsection{Interactions in the hexagonal lattice models}

The real-space dielectric function is useful in connection with the
``fractional point charge approximation'' introduced in
\cite{PhysRevX.8.031087}. Within this approximation, each Wannier
orbital consists of three point charges of strength $e/3$ positioned
in the middle of three neighbouring AA-regions of the moire
lattice. In this section we will estimate the screened direct Coulomb
potential between these $e/3$ charges based on the screened
interaction taking into account the screening from the substrate
(i.e. the second lowest panel in Fig. \ref{WRLongRangeFigure}).

\begin{figure}[t]
\includegraphics[width=\columnwidth]{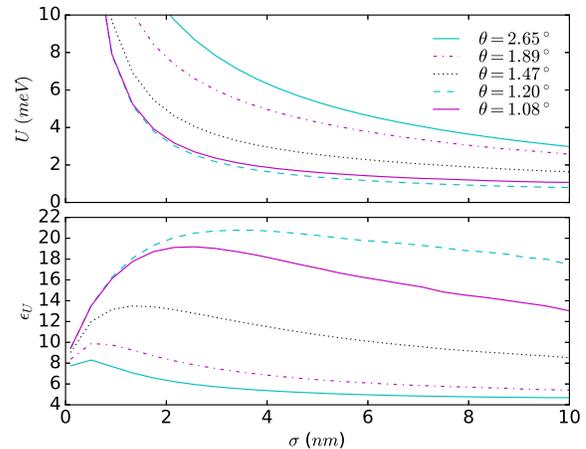}
\caption{Upper panel: The screened potential between two gaussian
  charge distributions of total charge $e/3$ and width $\sigma$. This
  approximates the on-site contribution of the $e/3$ fractional charge
  approximation \cite{PhysRevX.8.031087}.  Lower panel: The ratio of
  the screened potential and the bare potential, giving the effective
  dielectric constant of the on-site contribution for the fractional
  charges.\label{fractional_charge_onsite_figure}}
\end{figure}

\begin{figure}[t]
\includegraphics[width=\columnwidth]{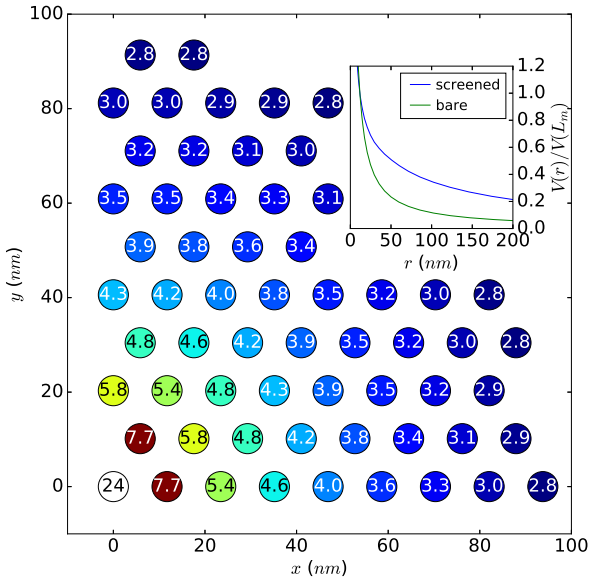}
\caption{Main panel: The potential between the $e/3$ charges in the
  AA-regions of the moire lattice in units of $0.1 \: meV$. Inset:
  Comparison of the decay of the bare and the screened
  interaction. \label{fractional_charge_long_range_figure}}
\end{figure}

To estimate the ``on-site term'' $U$ for the $e/3$ charges we
calculate the potential between two identical gaussian charge
distributions with different variances $\sigma$ and plot the result in
Fig. \ref{fractional_charge_onsite_figure}. We also calculate an
effective dielectric constant $\epsilon_U$ for the on-site
contribution as the ratio of the screened and unscreened
potentials. This yields values in the range $18$ to $22$ for a wide
range of $\sigma$.
Using numerical data for the Wannier functions and assuming a constant
dielectric function $\epsilon$, the on-site interaction $U$ was
evaluated in \cite{PhysRevX.8.031087} to be
$U \approx (e/3)^2/(0.28 \epsilon L_m)$. Here we take the value of the
dielectric function $\epsilon_U$ from
Fig. \ref{fractional_charge_onsite_figure} at $\sigma=0.28 L_m$, and
then plug in the moire lattice constant $L_m$ and the effective
dielectric constant $\epsilon=4 \pi \epsilon_0 \epsilon_U$ to the
above expression for $U$. Directly reading off $U$ from the upper
panel of Fig. \ref{fractional_charge_onsite_figure} at
$\sigma=0.28 L_m$ yields similar results.
For the rest of the potentials we use the point charge approximation
\cite{PhysRevX.8.031087} and directly use the value of the real-space
dielectric function for the distance between the point charges. The
results for the magic angle are plotted in
Fig. \ref{fractional_charge_long_range_figure}. It is interesting to
observe that the screened interaction decays \emph{slower} than $1/r$
for intermediate length scales, as demonstrated in the inset, which is
a feature of the 2D screened interaction,
Eqn. \ref{ScreenedCoulomb2D}.

We end this section by tabulating the Coulomb interaction terms for
the full Wannier orbitals in the magic angle case up to the fourth
nearest neighbour in Table \ref{lattice_model_V_table}, which can
easily be evaluated using
Fig. \ref{fractional_charge_long_range_figure}. We have also similarly
evaluated the interaction parameters for two other twist angles with
the corresponding values of $L_m$ and $\epsilon_U$. For the magic
angle case the short-range interaction terms are about $10 \: meV$,
which is roughly the same as the total bandwidth of the four
low-energy bands, and the Coulomb effects are thus expected to be
important. However, as mentioned in the introduction section, it could
be that the actual bandwidth at the angle $\theta \approx 1.1 \: \deg$
used in the transport experiments
\cite{CaoSuperconductivity,CaoInsulator,Yankowitz1059,2019arXiv190205151C,Sharpeeaaw3780}
is considerably larger, several tens of $meV$
\cite{Kerelsky_2019,2019arXiv190102997C,PhysRevB.99.201408}. Nevertheless,
the Coulomb interaction is of the same order of magnitude as the
experimentally measured width of the van Hove singularities
\cite{Kerelsky_2019}.
Our estimate of the on-site interaction $V_0$ of the full Wannier
orbitals is a bit smaller than the estimates of the on-site
interaction given in \cite{2019arXiv190411765P}. The differences are
perhaps partly because our dielectric function attains slightly larger
values than that in reference \cite{2019arXiv190411765P}, but also
because we use the fractional charge picture of the Wannier functions,
while reference \cite{2019arXiv190411765P} uses a momentum space
integral.

\begin{table}
\begin{tabular}{c | c | c || c | c | c | c | c}
$\theta$ & $\epsilon_U$ & $U$ & $V_0$ & $V_1$ & $V_2$ & $V_3$ & $V_4$\\
\hline
$1.47 \deg$ & $13$ & $4.6$ & $25$ & $22$ & $19$ & $18$ & $14$ \\
$1.20 \deg$ & $20$ & $2.4$ & $12$ & $10$ & $8.0$ & $7.7$ & $5.5$ \\
$1.08 \deg$ & $19$ & $2.3$ & $13$ & $12$ & $9.8$ & $9.5$ & $7.6$ \\
\end{tabular}
\caption{ Estimated $n$:th nearest neighbour interactions $V_n$
  (energy unit $meV$) for the full Wannier orbitals for the magic
  angle case $\theta=1.20$ and for TBG slightly above and below the
  magic angle. This can be compared with Table I in
  \cite{PhysRevX.8.031087}. U here denotes the ``on-site'' term of
  the $e/3$ charges, while $V_0$ is the on-site term for the full
  Wannier orbitals.
  \label{lattice_model_V_table}
}
\end{table}

\subsection{Short range contributions}

In the previous section we discussed the screening of the long-range
part of the interaction. To complete the treatment we also need to
estimate the remaining short-range contribution. To ensure a hermitian
interaction matrix we write the screened interaction in the form
\begin{equation}
\begin{split}
W(\mathbf{q}) &= \epsilon(\mathbf{q})^{-1} V(\mathbf{q}) \\
& = \frac{1}{2} \left( \epsilon(\mathbf{q})^{-1} V(\mathbf{q}) + V(\mathbf{q}) (\epsilon(\mathbf{q})^\dagger)^{-1} \right).
\end{split}
\end{equation}
The short range part $W_s(\mathbf{q})$ of the interaction is then
obtained by replacing $V$ on the second line by the bare short-range
interaction. As the short-range interaction is insensitive to the
small-$q$ details of the polarization function, we use here simply the
RPA dielectric function of uncoupled graphene layers. We do not
consider here the screening from the $\sigma$-bands or the
substrate. We take the Fourier transform of the bare intralayer
interaction numerically, and use the continuum approximation
\ref{LongRangeVEquation} for the bare interlayer interaction.

\begin{figure}[t]
\includegraphics[width=\columnwidth]{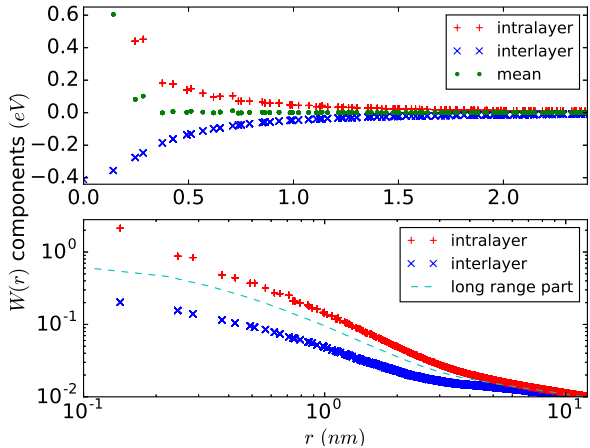}
\caption{Upper panel: Intralayer and interlayer components of the
  short-range part of the screened interaction as a function of the
  planar distance. The mean of the two components is also shown. Lower
  panel: The full intralayer and interlayer interaction for
  $\theta=1.2$ calculated as the sum of the long-range part and the
  short-range components. Here only the screening from the moire bands
  has been taken into account. \label{short_range_real_space_W}
}
\end{figure}

The screened short-range part of the interaction is plotted in
Fig. \ref{short_range_real_space_W}. We note that, although the
\emph{bare} short range interaction defined in
Eq. \ref{ShortRangeInteractionDefinition} is purely intralayer, the
\emph{screened} interaction has an interlayer component due to the
interlayer parts of the dielectric matrix, where the full interaction
still enters. The interlayer part of the screened short-range
interaction is negative, which compensates for some screening missing
from the long-range part, but the total Coulomb interaction is still
always positive. The intralayer part is positive and compensates for
the ``artificial'' short-range screening $\exp(-q d_{il})$ introduced
to the long-range part.

The short-range part produces non-negligible corrections to intra- and
interlayer interactions even beyond $r \sim 10 \: nm$. However, the
mean of the interlayer and intralayer corrections is quite small
already for $r \gtrsim 0.25 \: nm$, corresponding to the next-nearest
neighbour interaction within the graphene layer. Thus the long-range
part of the interaction should be a reasonably good approximation for
everything but the on-site and nearest neighbour interactions provided
that the charge distribution is equal in both layers. The correction
for the intralayer on-site term of the microscopic model is roughly
$5 \: eV$. Within the fractional charge picture the $e/3$ charge is
distributed to one AA-region of the unit cell. Estimating that this
region contains of the order of $1000$ atoms and the charge is equally
distributed over each atom, the on-site terms contribute an energy of
$0.5 \: meV$. Thus the contribution from the short-range part to the
on-site term of the $e/3$ charges is likely to be less than
$1 \: meV$. Using similar arguments it was concluded in
\cite{Guinea13174} that the microscopic on-site contribution is
relatively unimportant. Taking into account the distance dependent
screening from the moire bands increases the importance of the short
range contributions because of the stronger screening at intermediate
distances compared to very short ones. This can be taken into account
by slightly reducing $\epsilon_U$ (see table
\ref{lattice_model_V_table}).

\section{Discussion and Conclusions}

An accurate low-energy theory of twisted bilayer graphene based on
microscopic tight-binding models seems to be a realistic goal,
although a more complete understanding of e.g. lattice relaxation
effects is still needed. The simplest models have been built on the
four lowest moire bands
\cite{PhysRevX.8.031087,PhysRevX.8.031088,PhysRevX.8.031089}. Here we
have discussed screening of the Coulomb interaction by the moire bands
outside this model space. As expected for a two-dimensional gapped
system, the cRPA polarization function is quadratic for small momentum
exchange with a 2D polarizability $\alpha_{2D}$ depending on the twist
angle. The polarizability attains a peak value of
$\alpha_{2D} \approx 100 \: nm$ at the magic angle, and is of the
order of tens of nanometers when the angle is tuned $\sim 0.3$ degrees
away from the peak. The quadratic approximation is valid up to a
momentum scale that is expected to follow $q \sim \Delta / v_F$, where
$\Delta$ is the distance from the Fermi level to the lowest screening
bands. In freestanding TBG the 2D screened Coulomb interaction
\cite{PhysRevB.84.085406} is then a reasonable approximation for
$r \gtrsim v_F/\Delta$. At the magic angle with
$\Delta \sim 10 \: meV$ this is a relatively large distance,
$r \sim 50 \: nm$. However, as the experimentally observed bandwidth
and $\Delta$ e.g. at the angle $\theta \approx 1.1 \deg$ seem
significantly larger
\cite{Kerelsky_2019,2019arXiv190102997C,PhysRevB.99.201408}, this
approximation might be sufficient for length scales somewhat larger
than the moire lattice constant.

For large enough momenta the polarization and the dielectric function
of freestanding TBG become independent of the twist angle and start to
approach that of unhybridized graphene layers. The momentum scale for
this transition is expected to depend on the strength $V_{il}$ of the
interlayer hopping parameters as $q \sim V_{il}/v_F \approx 1 \: nm$.
In the intermediate region between the low- and high-momentum limits
the dielectric function has a maximum which is largest for twist
angles close to the magic angle, attaining values of
$\epsilon(q_{max}) \approx 25$. Screening from the moire bands
dominates effects from the graphene $\sigma$-bands, although a more
accurate estimation of $\sigma$-band effects at short distances would
require density functional theory calculations.

While dealing with the full tight-binding model is computationally
costly, it is also flexible and can easily be adapted to changes in
the tight-binding parameters, for example. Preliminary results
indicate that doping the system by two electrons per moire cell does
not change the essential features of the polarization function. An
interesting experimental development is increasing the interlayer
coupling with pressure \cite{Yankowitz1059}, which would mean
increasing the local interlayer hopping scale $V_{il}$. Increased
pressure is thus expected to enlarge the region where the polarization
function shows strong dependence on the twist angle.

Screening from the TBG environment includes contributions from the
substrate, typically boron nitride, and metallic screening from the
gate electrodes.  The effect of the boron nitride layers is mainly to
renormalize the long-range interaction, while most of the momentum
dependence of the dielectric function still arises from the intrinsic
TBG screening. In contrast, metallic screening from the electrodes
will provide an interaction cutoff distance proportional to their
distance from the TBG. This provides interesting possibilities for
experiments: As a result of the 2D screening behaviour, the long-range
interaction without the metallic screening decays slowly, even slower
than $1/r$. A device with a large separation of the electrodes from
the TBG is thus expected to have strong extended interactions, while a
small separation leads to more local interactions. A comparison of the
phase diagrams for such devices might give clues about the mechanism
of the superconductivity. For instance, it is theoretically expected
that the $d$-wave superconductivity in the square lattice Hubbard
model is suppressed by nearest neighbour repulsion
\cite{PhysRevB.97.184507}, although retardation effects could reverse
this picture \cite{PhysRevB.94.155146}.

Our results can be compared to the continuum model calculations of
reference \cite{2019arXiv190411765P}, where especially the tuning of
the short-range interaction by changing the dielectric environment is
discussed. There the cRPA screened on-site interaction in the
magic-angle case was found to be $40 \: meV$ for the freestanding TBG,
while a metallic gate $3 \: nm$ from the TBG reduces this to
$28 \: meV$. Thus the gate has to be relatively close to significantly
alter the on-site interaction, and at larger gate distances the main
effect is on the long-range part. Quantitatively our estimates for the
on-site contribution in the freestanding case are a bit smaller than
in \cite{2019arXiv190411765P}, $12 \: meV$ to $25 \: meV$ within
$\pm 0.3 \deg$ of the magic angle. While the calculations in
\cite{2019arXiv190411765P} and our work provide a reasonable picture
of the screening, improved microscopic models with the magic angles
and bandwidths coinciding with the experimentally measured ones are
still needed to truly fix the parameters of the low-energy model. It
has also been shown that the cRPA can overestimate the screening in
some models \cite{PhysRevB.98.155132,PhysRevB.98.235151}, especially
at short distances. Indeed, a more accurate treatment of the
short-range interactions free of the momentum-diagonal approximation
made in our work would be desirable.

Assuming that the superconductivity and insulating states close to
half-filling of the conduction and valence bands result from strong
correlation effects, the scale of the interaction should roughly
correspond to the width of the van Hove singularities, which has been
measured to be of the order of $10 \: meV$ \cite{Kerelsky_2019}. This
is indeed the scale of the interactions produced by the cRPA
calculation, and thus our results are compatible with a picture where
the screened Coulomb interaction is the driving force behind the
observed insulating states. As noted e.g. in
\cite{2019arXiv190411765P}, one thing that speaks against this picture
is the very small thermal activation gap of $0.3 \: meV$ in the
insulating states at half-filling of the valence or conduction bands
\cite{CaoInsulator}, as compared to the effective interaction
strength, although spectroscopy experiments indicate larger gaps by an
order of magnitude \cite{Kerelsky_2019}, and it has been suggested
that the small transport gaps could be due to disorder averaging
\cite{Kerelsky_2019}.

Finally, we consider TBG at the charge neutrality point. A
long-standing theoretical problem has been the stability of Dirac
liquids under the Coulomb interaction
\cite{RevModPhys.81.109,PhysRevLett.113.105502,PhysRevLett.118.026403,Tang570},
and it would be interesting to study how this problem is modified by
the 2D screening and the doubling of the number of components as
compared to single-layer graphene. It has been proposed that TBG at
charge neutrality is close to a quantum critical point where a Dirac
mass term is spontaneously generated \cite{2019arXiv190111424D}. As
the extended interaction terms play an important role for such phase
transitions in Dirac systems \cite{2019arXiv190111424D,Tang570}, the
possibility to tune them experimentally is interesting also in this
context.

\begin{acknowledgments}
  This project was supported by FP7/ERC Consolidator Grant QSIMCORR, No. 771891, and the Deutsche Forschungsgemeinschaft (DFG, German Research Foundation) under Germany's Excellence Strategy -- EXC--2111--390814868. T.I.V. acknowledges funding from the Alexander von Humboldt foundation.
\end{acknowledgments}

\vspace{5 mm}

\emph{Note added: } After the completion of this work, we became aware
of a recent preprint that also considers the cRPA polarization
function of twisted bilayer graphene \cite{2019arXiv190900591G}.

\bibliography{combined_references.bib}{}

\end{document}